\documentclass[aps,prb,twocolumn,citeautoscript,superscriptaddress,fleqn]{revtex4}  
\usepackage[english, german, french]{babel}
\usepackage[latin1]{inputenc} 

\usepackage{amsmath}
\usepackage{amssymb}
\usepackage{graphicx}  
\usepackage{array}
\usepackage{color}
\usepackage{wasysym}
\usepackage{courier}
\usepackage{hyperref}
\usepackage{siunitx}

\begin{document}

\title{Quick X-ray microtomography using a laser-driven betatron source}

\begin{abstract}
Laser-driven X-ray sources are an emerging alternative to conventional X-ray tubes and synchrotron sources. We present results on microtomographic X-ray imaging of a cancellous human bone sample using synchrotron-like betatron radiation. The source is driven by a 100-TW-class titanium-sapphire laser system and delivers over $10^8$ X-ray photons per second. Compared to earlier studies, the acquisition time for an entire tomographic dataset has been reduced by more than an order of magnitude. Additionally, the reconstruction quality benefits from the use of statistical iterative reconstruction techniques. Depending on the desired resolution, tomographies are thereby acquired within minutes, which is an important milestone towards real-life applications of laser-plasma X-ray sources.
\end{abstract}

\author{A. Döpp}
\affiliation{Ludwig-Maximilians-Universitat München, Am Coulombwall 1, 85748 Garching, Germany}

\author{L. Hehn}
\affiliation{Lehrstuhl für Biomedizinische Physik, Physik-Department \& Munich School of BioEngineering, Technische Universität München, Garching 85748, Germany.}

\author{J. Götzfried}
\affiliation{Ludwig-Maximilians-Universitat München, Am Coulombwall 1, 85748 Garching, Germany}

\author{J. Wenz}
\affiliation{Ludwig-Maximilians-Universitat München, Am Coulombwall 1, 85748 Garching, Germany}

\author{M. Gilljohann}
\affiliation{Ludwig-Maximilians-Universitat München, Am Coulombwall 1, 85748 Garching, Germany}

\author{ H. Ding}
\affiliation{Ludwig-Maximilians-Universitat München, Am Coulombwall 1, 85748 Garching, Germany}

\author{S. Schindler}
\affiliation{Ludwig-Maximilians-Universitat München, Am Coulombwall 1, 85748 Garching, Germany}

\author{F. Pfeiffer}

\email{franz.pfeiffer@tum.de}
\affiliation{Lehrstuhl für Biomedizinische Physik, Physik-Department \& Munich School of BioEngineering, Technische Universität München, Garching 85748, Germany.}
\affiliation{Institut für Diagnostische und Interventionelle Radiographie, Klinikum rechts der Isar, Technische Universität München, 81675 München, Germany.}

\author{ S. Karsch}
\email{stefan.karsch@mpq.mpg.de}
\affiliation{Ludwig-Maximilians-Universitat München, Am Coulombwall 1, 85748 Garching, Germany}
\affiliation{Max Planck Institut für Quantenoptik, Hans-Kopfermann-Str. 1, Garching 85748, Germany}

\maketitle

Radiography is the oldest and arguably most prevalent application of X-rays. The X-rays used for this application are usually produced in X-ray tubes, where electrons from a thermo-cathode hit an anode, leading to bremsstrahlung and line emission. This design has been established by William D. Coolidge and contemporaries \cite{Coolidge:1945hl} and has since then remained essentially unchanged. In parallel to the development of X-ray tubes, accelerator-based synchrotron sources and, most lately, X-ray Free-Electron-Lasers have been developed\cite{McNeil:2010dp}. While these sources allow the production of X-rays at unrivaled brightness, they have never been available for a wide user base due to their immense costs and size.

In response to this, in recent years, novel X-ray sources based on high-power laser systems are receiving increased attention. As lasers can create high fields over short time scales, these sources provide alternative means for radiation production. While some schemes closely resemble conventional X-ray tubes in terms of geometry and function \cite{Rousse:1994wh,Dopp:2016faa}, other approaches rely on entirely different physical phenomena, e.g. high-harmonic generation in gases\cite{Popmintchev:2012gh}. Among the most promising sources are all-optical light sources, which miniaturize the principle of a modern synchrotron\cite{Corde:2013bja}. Here, a laser is focused on a gaseous target and creates, in its wake, a plasma wave capable of accelerating electrons at fields of gigavolt to teravolt per meter\cite{Esarey:2009ks}. Due to the similarly strong radial fields, electrons with initial transverse momentum will perform betatron oscillations that lead to wiggler-like, broadband X-ray emission\cite{Rousse:2004tc}. Alternatively, a laser pulse can be used to provoke radiation emission via Thomson scattering\cite{Khrennikov:2015gxa}. In contrast to X-ray tubes, these sources provide collimated X-ray beams at a divergence of several milliradians and ultrashort pulse durations.

Relying on a tightly focused laser pulse, all of these sources have their small, micrometer source sizes in common; hence the potential of laser-based sources for high-resolution X-ray imaging has been discussed since the 1990s \cite{Krol:1997ec}. To date, X-ray imaging has been demonstrated for k-alpha \cite{Toth:2005ei}, betatron \cite{Fourmaux:2011cs} and Compton sources\cite{Dopp:2016ki}. Betatron sources have received particular attention due to their high photon flux, and first proof-of-principle experiments demonstrated tomographic reconstructions  \cite{Wenz:2015if,Cole:2016fh}. Radiographs from betatron radiation also commonly show strong edge enhancement caused by Fresnel diffraction\cite{Kneip:2011cx}. In a single-material scenario, these edge-enhanced images can be related to phase maps using the transport-of-intensity equation\cite{Paganin:2002hp}. Based on this method, we have recently demonstrated the first quantitative tomographic reconstruction of absolute electron densities of a lacewing insect \cite{Wenz:2015if}. In this letter, we present measurements at increased repetition rate, photon energy and photon yield. Furthermore, we extend our measurements to samples of more eminent interest for radiological applications.

The measurements were performed at the Laboratory for Extreme Photonics in Garching, Germany, using the ATLAS titanium-sapphire (Ti:Sa) laser system. Including beam transport, this standard CPA laser delivers pulses of $1.9\pm0.1$ J energy and 27 fs duration, resulting in a peak power of 65-75 TW on target. An off-axis parabolic mirror is used to focus the laser to a spot size of \SI{30}{\micro\metre} (FWHM), where it reaches a peak intensity of $5.5\times10^{18}$ W/cm$^2$.

\begin{figure}[bt]
\includegraphics[width=.9\linewidth]{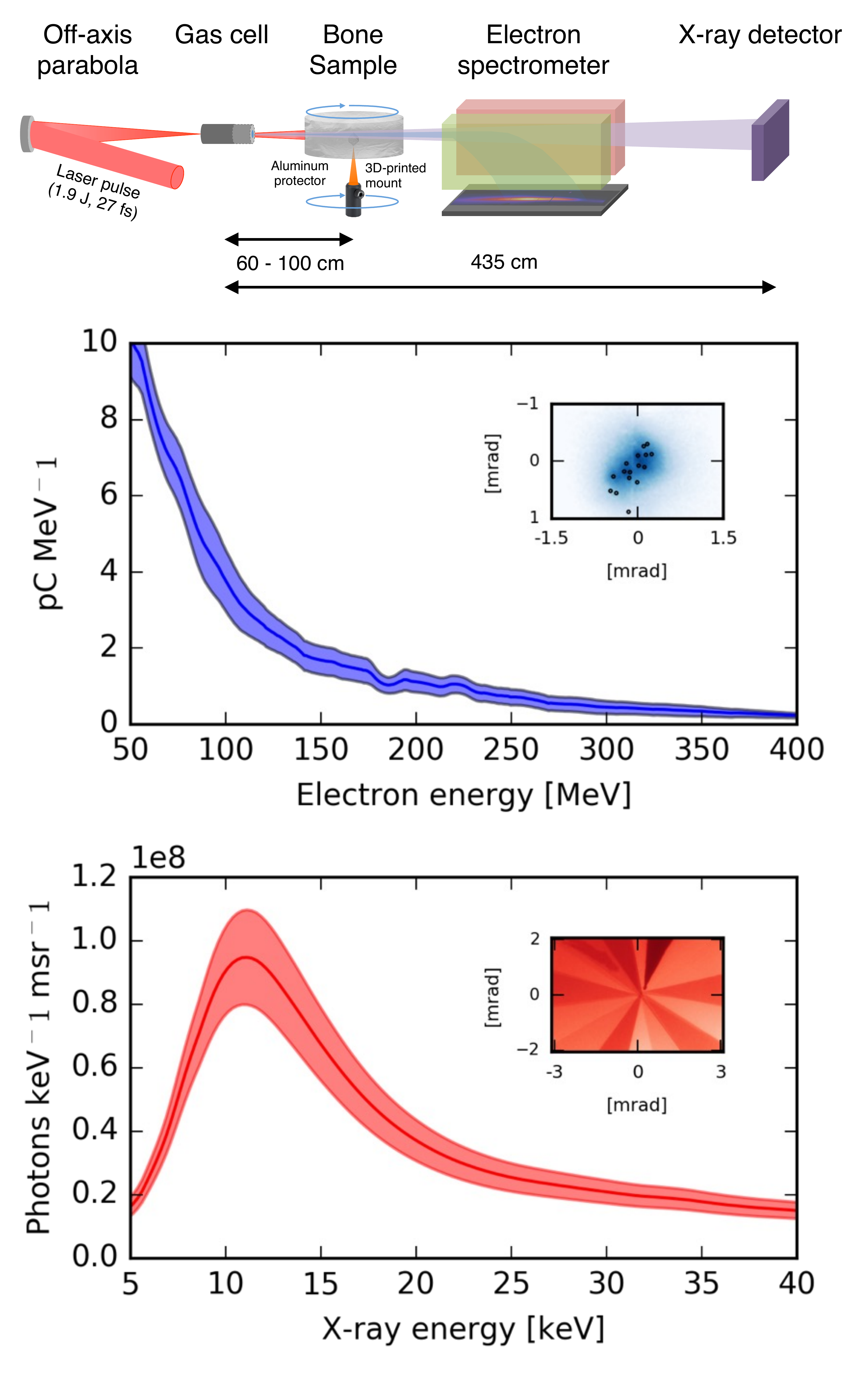}

\caption{Top: Experimental layout (not to scale). The laser pulse is focused using an f/25 off-axis parabola on a hydrogen gas cell of variable length. As a result of the laser-plasma interaction in the cell, electrons and X-rays are generated. The sample is placed between the gas cell and the dipole magnet spectrometer, protected by an aluminum foil. X-rays are detected on the X-ray detector with a geometrical magnification of $\sim$4-7. Middle: Average electron beam spectrum and beam profile (inset) for 20 consecutive shots. The shaded area indicates the RMS error; dots in the inset show the beam pointing of individual shots. Bottom: Average X-ray spectrum at the detector as reconstructed from filter transmissions over 20 consecutive shots. The shaded area indicates the RMS error; the inset shows the average image of the filter transmissions.}
\label{foto}
\end{figure}

To create an underdense plasma as medium for wakefield acceleration, a hydrogen gas cell of variable length (5-15 mm) is placed in focus. Plasma electrons are self-trapped in the laser's wakefield and accelerated. To determine their energy, a dipole magnet (0.8 m, 0.85 T) is placed 1.8 m from the source. The magnet deflects electrons onto a scintillating screen, which is imaged using a camera system. Absolute beam charges are obtained by comparing the scintillation light from electrons hitting the screen to a calibrated tritium capsule \cite{Buck:2010bt}. While the electron energy on the spectrometer may exceed 500 MeV energy for short gas cell lengths (not shown here), the X-ray yield is optimized for long propagation length ($10-13$ mm). In this case, the resulting spectrum has an exponential shape, and the final electron energy is lower due to dephasing, but electrons can perform more betatron oscillations and therefore radiate stronger. The long acceleration also contributes also to stabilizing the X-ray energy (see below). By optimizing the gas density ($n_e\sim 5\times 10^{18}$ cm$^{-3}$) and the laser focus, we reached an electron beam charge of $736\pm51$ pC. The electron beam pointing is determined by the injection process and the non-linear laser propagation in the plasma; it varies by $1.1\pm0.1$ mrad.

To characterize the radiation, we have used two different detectors. First, we used a direct detection X-ray CCD camera (Princeton PIXIS-XO), which has a known quantum efficiency. While such cameras can be used to determine the X-ray energy using single photon counting, in our case the peak X-ray flux at the detector has been too high for this kind of measurement. Instead, a set of different aluminum filters is placed in front of the detector (see inset in Fig.1). Based on the transmission ratio of each filter, the X-ray energy is estimated using an iterative algorithm \cite{Sidky:2005eh}. An averaged spectrum of 20 shots is shown in Fig.1. Note that this is the spectrum reaching the detector and low energy X-rays are filtered by a 15 \SI{}{\micro\metre}-thick aluminum filter and 125  \SI{}{\micro\metre} Kapton foil. 

Based on the spectral measurements, we calculate an average photon flux of $(1.6\pm0.35)\times10^{9}$ photons msr$^{-1}$ s$^{-1}$ at 1 Hz repetition rate. The total flux has therefore increased compared to earlier experiments (including the different filter configuration), while the RMS fluctuations are similar \cite{Wenz:2015if}. As the spectrum is synchrotron-like, it can also be characterized by a critical energy $\hbar\omega_{cr}=13.5 \pm 0.95$ keV. The spectral shape is therefore relatively stable ($\pm7\%$), which can be explained by our experimental setup: The detected X-ray spectrum is the sum of radiation emitted over the entire accelerator length. By extending the accelerator length far beyond dephasing, we are effectively smoothing the spectrum and thus reducing shot-to-shot fluctuations. In consequence, a higher energy stability is achieved, which is essential for tomographic reconstruction.

X-ray CCD cameras offer a high spatial resolution and were used in preceding studies for propagation-based phase contrast imaging of small insects \cite{Fourmaux:2011cs,Wenz:2015if}. But since the quantum efficiency of the CCDs decreases with the X-ray energy, these detectors are unsuitable for imaging of denser objects like human bone. Here we have used a Gd$_2$O$_2$S:Tb based scintillator (P43) as a substitute, which is fiber-coupled to a $1388\times1038$ pixel CCD camera. The camera uses a 50:11 taper, where each pixel represents an area of $(29$ \SI{}{\micro\metre}$)^2$ on the scintillator screen. Placed at 435 cm from the source the field of view covers approximately $ 9.2\times 6.9 $ mrad$^2$. For most shots the betatron beam divergence is large compared to the field of view, but for some shots a bi-gaussian beam profile can be fitted, giving a lower estimate of $11.6 \times 6.0$ mrad. The total measured photon yield is therefore of the order of $>10^8$ photons per shot. 

\begin{figure*}[bt]
\includegraphics[width=1.0\linewidth]{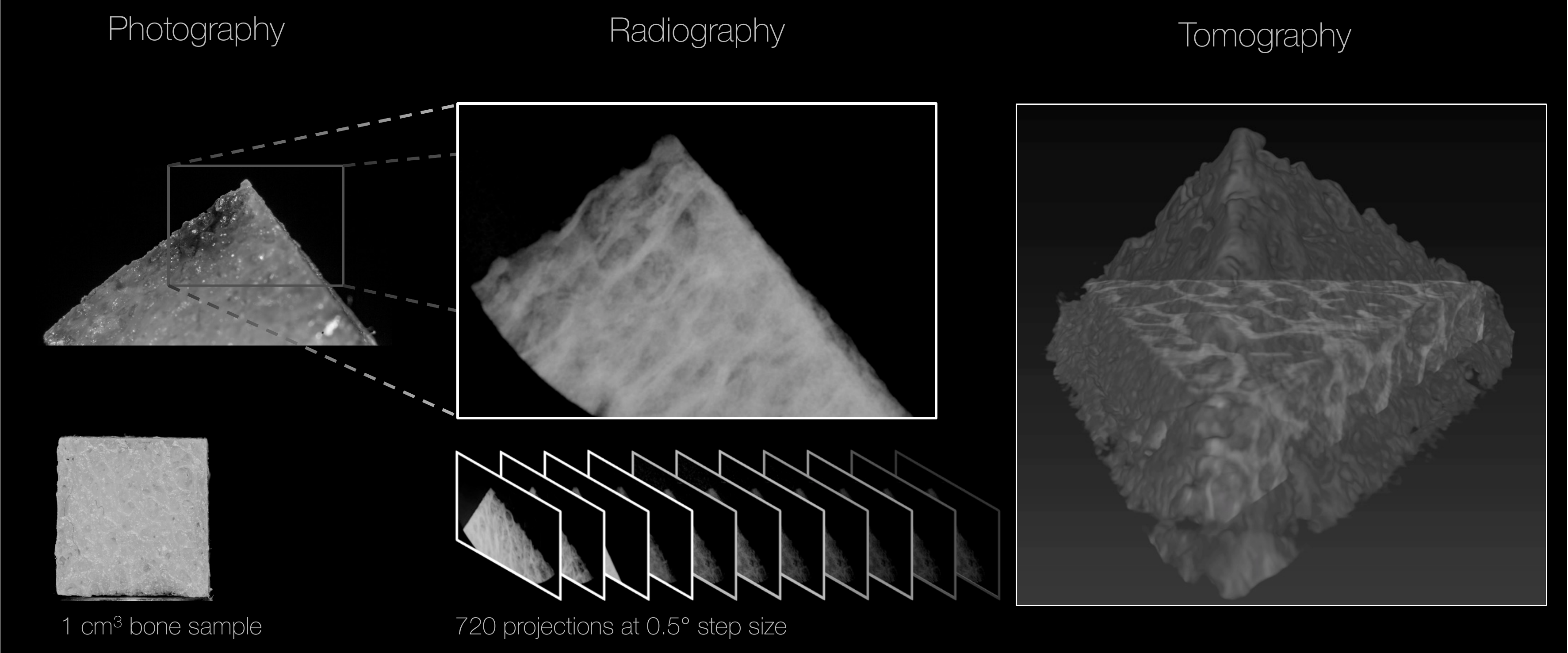}

\caption{Tomography of cancellous bone sample. Bottom left: Photograph of the sample. Top left: Macro photography as in the experimental setup. Center: Radiograph using the scintillator camera. Right: Rendering of the tomographic reconstruction with statistical image reconstruction (SIR), showing transverse and longitudinal cross sections.}
\label{foto}
\end{figure*}

While the laser-plasma source can operate at 5 Hz (determined by the pump lasers of the last amplifier), the acquisition time in this experiment was limited by the readout time of the detector and measurements were performed at 0.5 - 1.0 Hz. Compared to 0.025 - 0.1 Hz for earlier measurements \cite{Wenz:2015if,Cole:2016fh}, this leads to a significantly reduced acquisition time for a tomography scan. 
To demonstrate the source's potential for medical imaging, we use a sample of cancellous human bone tissue, which contains fine lattice structures (see Fig.2). These \textit{trabeculae} are affected by osteoporosis and understanding their degradation is a typical application of $\mu$CT in biomedical research \cite{Stock:2008wc}. 

As shown in Fig.1, the sample is mounted at 65 to 100 cm distance from the source. It is protected from laser radiation using \SI{15}{\micro\metre} thick aluminum foil and a static fiducial is placed next to the rotating bone sample. First, we have acquired projections in the full 360 degree range in steps of 0.5 degree. For the first 90 degrees, 10 images were acquired at each position, while the remaining data contains a single image per angle. In order to study the image quality enhancement as function of the number of shots per angle, we furthermore acquired $>100$ shots for 10 different angles. 

In post-processing, we employ a flat-field correction and then use normalized cross correlation on the fiducial for image registration. While the latter is necessary to compensate for pointing fluctuations of the X-ray beam, it can simultaneously be used to achieve sub-pixel resolution images \cite{Wenz:2015if}. After registration, all images from the same angle are summed. We find that for our setup, the influence of flux fluctuations rapidly decreases for $>3$ images per angle and the image quality does not improve noticeably for $>20$ shots per position. The registered projections of the entire dataset are then summed and the rotation axis is determined. To account for slight variations in the illumination on the detector, the beam-profile is fitted with a two-dimensional ramp from the sample-free regions.

\begin{figure}[b]
\includegraphics[width=1.\linewidth]{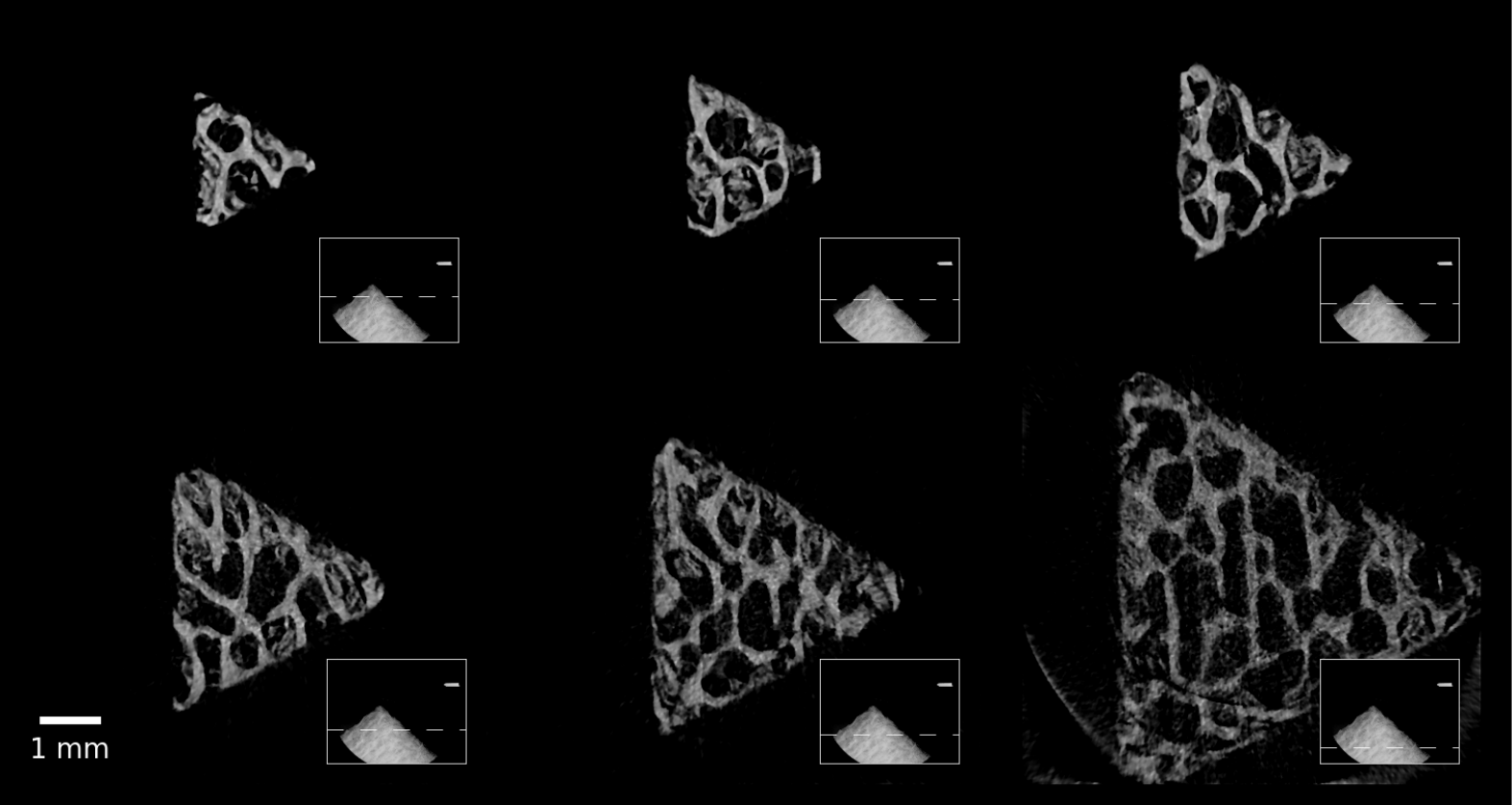}

\caption{Projection slices for the tomographic reconstruction with SIR. Insets show the position of the slices relative to the raw data.}
\label{foto}
\end{figure}

Due to the polychromaticity of the source, the line-integrals are underestimated due to beam-hardening. This leads to lower attenuation coefficients in the reconstructed volume at regions where the sample is thick. To account for this effect, the line-integrals are heuristically corrected by an additional factor, depending on the thickness of the sample.

For the subsequent tomographic reconstruction, a statistical image reconstruction (SIR) technique \cite{Fessler:2000tm} is used that proves to provide improved image quality compared to conventional filtered back-projection techniques as discussed later. The corresponding penalized-likelihood objective function that is used in the following can be written as
$
\hat{\mu} = \arg\min ( A\mu - \hat{l} )_W^2 + \lambda R_\gamma(\mu).
$
Thereby, the vector of the object attenuation coefficients is denoted by $\mu$. The linear forward projection operation is modeled by a matrix $A$. Due to the small angular size occupied by the sample, the transverse displacement of photons is of the order of the detector resolution and the projections are assumed parallel. The estimated line-integrals that are recovered from the post-processed projections are denoted by $\hat{l}$ and $W$ is a statistically motivated weighting factor, which is set to zero for detector elements that are not illuminated. As regularization a Huber \cite{Huber:2011uf} penalty $R_\gamma$ is used that penalizes small value differences between neighboring voxels quadratically and large ones linearly. The transition point $\gamma$ between the two cases is chosen according to the noise level in the reconstructed volume. The strength of the regularization is denoted by $\lambda$ and has to be set by hand. 

Figure 3 illustrates transverse slices of the reconstructed volume. With a voxel size of $($\SI{6.7}{\micro\metre}$)^3$, the resolution is comparable to conventional $\mu$CT on the top slices\cite{Peyrin:1998tj,Cooper:2007bl}. At the bottom the tomographic dataset is not complete, due to the limited field of view (cf. Fig. 3 bottom right). Nevertheless, the SIR can produce reconstructions for these slices, although at reduced image quality.  

The tomographic scans from the preceding part were performed over the course of one hour and covered 720 different angles with multiple exposures per projection angle. To explore the limitations of the technology we removed the fiducial, moved the sample closer to the source (65 cm) and performed another tomography run covering 180 degrees at 1 degree step size and a single exposure per step. Using 1 Hz repetition rate, the entire tomography was acquired in three minutes.

\begin{figure}[t]
\includegraphics[width=1.\linewidth]{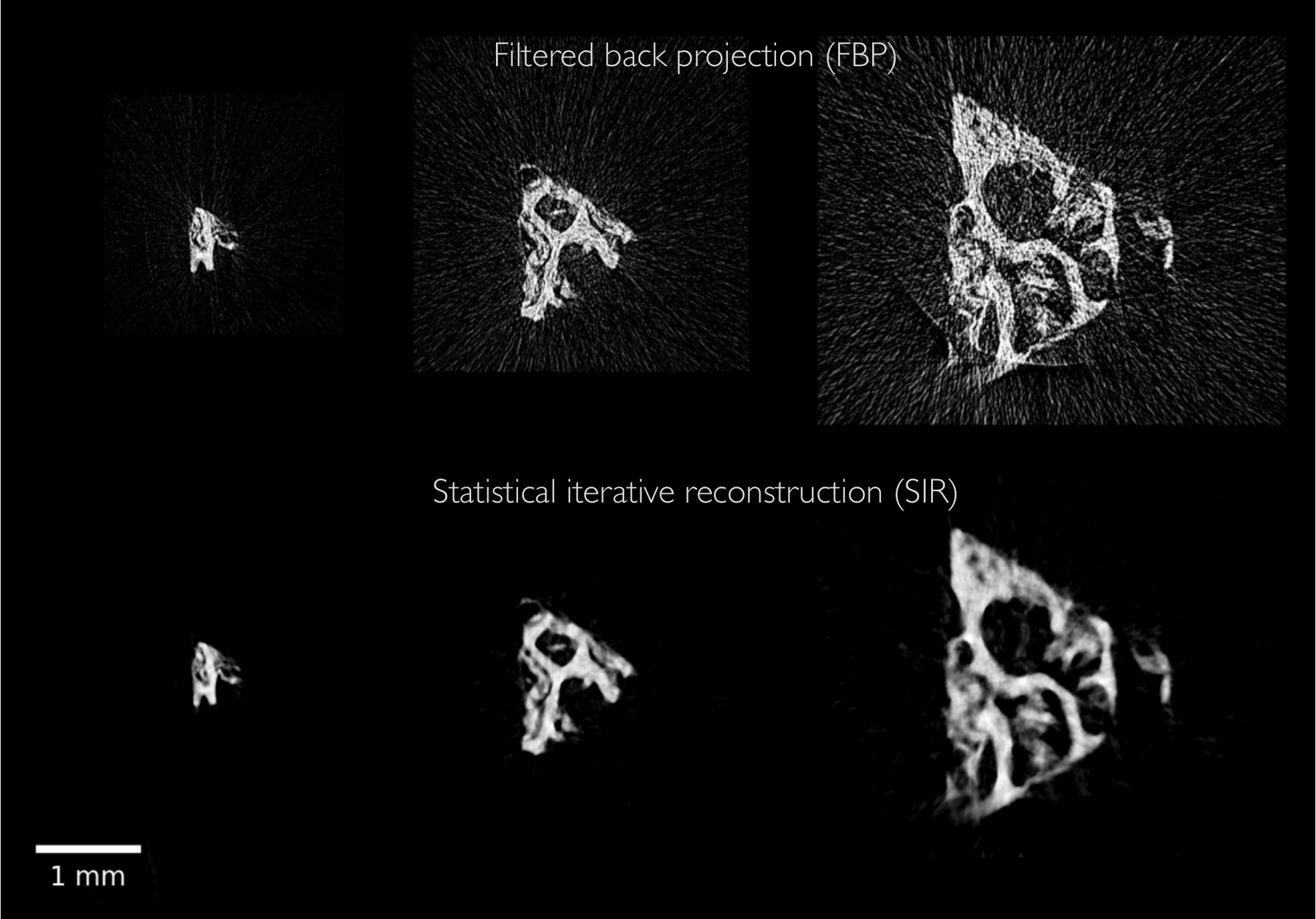}

\caption{Quick tomography: Comparison between reconstruction with filtered back-projection (top) and statistical iterative reconstruction (bottom).}
\label{foto}
\end{figure}

As the fiducial is omitted, the previous alignment method using normalized cross-correlation cannot be applied anymore. To overcome this limitation an iterative alignment scheme is used. Thereby, we exploit the fact that a filtered back-projection of the estimated line-integrals followed by a forward-projection does only recover the estimated line-integrals if they are consistent with its 3D representation. The relative displacement between the re-projected and estimated line-integrals is used to align the projections\cite{GuizarSicairos:2015eh}. All other post-processing steps are analogue to the previous measurement.

To highlight the benefits of SIR, we compared this technique to the conventional filtered back-projection approach in Fig. 4 for different slices. With the statistical weights, we can compensate the effects of the limited field of view that are even more pronounced in this 180 degree scenario. Due to the regularization, the noise level is distinctively reduced without deteriorating the resolution significantly. Most importantly, the analytical reconstruction procedure leads to under-sampling artifacts, due to the limited number of acquired projections, which SIR techniques are capable to handle. The resolution of this quick, single-shot tomography is lower than in the first tomography scan. This is mostly due to noise and flux fluctuations, which is reduced for multi-shot acquisition. We found that summing 2-3 shots per angle using the fiducial results in a good tradeoff between acquisition time, dose and resolution.

In conclusion, we have presented results on X-ray microtomography of a human bone sample using a laser-driven imaging system that operates at 1 Hz. The photon flux more than $10^{8}$ photons per shot is sufficient for single-shot acquisition of projections and tomographies were generated using statistical iterative reconstruction. Furthermore, the resolution can be improved (to sub-pixel level) when several shots are aligned using a fiducial or, alternatively, using tomographic consistency.

Being able to acquire tomographies of dense objects within minutes, our results represent a major step towards real-life applications of laser-driven X-ray sources: 100-TW-class Ti:Sa lasers, as used in this study, have been under constant  developement since the early 2000s\cite{Pittman:2002bs} and are now produced commercially. The betatron source itself reliably produces bright X-rays with high spatial coherence. In future experiments both the flux and fluctuations of the source could be further improved using ionization-induced injection in gas mixtures\cite{Dopp:2017dza}. Additionally, novel wiggler concepts like nano-wires\cite{Andriyash:2014ea} or hybrid wakefield accelerators\cite{Ferri:2017tia} could increase both energy and photon yield by an order of magnitude. The repetition rate and resolution are currently limited by the detector, an issue that can be resolved using high-resolution CMOS cameras. Commercial Ti:Sa laser technology would readily allow operation at up to 10 Hz and it was recently demonstrated that millijoule-class laser systems can be used for wakefield acceleration at kHz repetition rate\cite{Guenot:2017kv,Salehi:2017bp}. However, due to laser depletion, these sources do not achieve electron energies sufficient for betatron X-ray generation. Increasing the average power of femtosecond lasers beyond kilowatt is therefore an important challenge for the near future. For instance, coherent pulse combination has been demonstrated up to kilowatt average powers\cite{Muller:2016fz} and promises easy scalability to higher pulse energies\cite{Mourou:2013gl}. 

\newpage
\subsection*{Acknowledgements}

The authors thank F. Schaff and T. Baum (TUM) for providing bone samples. This work was supported by DFG through the Cluster of Excellence Munich-Centre for Advanced Photonics (MAP EXC 158), the DFG Gottfried Wilhelm Leibniz program, TR-18 funding schemes, by EURATOM-IPP and the Max-Planck-Society.

\end{document}